\documentclass[12pt]{article}				
\usepackage[tbtags]{amsmath}				
\def\d{{\rm d}}
\def\tr{{\rm Tr}}
\def\1/2{\frac{1}{2}}

\title{Constructive building of the Lax pair in the non-linear sigma model}

\author{Ariel O. Garc\'{\i}a\thanks{~E-mail: {\tt ariel@cab.cnea.edu.ar}} \
    and Roberto C. Trinchero\thanks{~Also supported by CONICET.
                             E-mail: {\tt trincher@cab.cnea.edu.ar}.} \\
        Centro At\'omico Bariloche and Instituto Balseiro \\
        (8400) Bariloche - RN \\
        Argentina}

\date{October, 1995}

\begin{document}

\begin{titlepage}
\maketitle

\abstract{A derivation of the Lax pair for the $(1+1)$-dimensional
non-linear $\sigma$-model is described. Its main benefit is to have a
clearer physical origin and to allow the study of a generalization to
higher dimensions.}

\vspace{1cm}

\vspace{1cm}
\noindent arch-ive/9511002

\end{titlepage}

\addtocounter{page}{1}

\section{Introduction}

It has been known for several years that the non-linear $\sigma$-model, in a
two dimensional space-time, has a non-local conservation law \cite{LuPo}
that can be cast as a non-local symmetry transformation \cite{EiFo,Do}. This
is usually shown to be true on the basis of the existence of a ``Lax pair'',
a pair of linear differential equations depending on a spectral parameter,
whose compatibility condition is exactly the field equation of the model.
Exact integrability of the system follows, via the inverse scattering
procedure \cite{dVe}.

An infinite set of non-local conserved charges can be directly obtained from
this Lax system, as in \cite{LuPo}, or as in \cite{EiFo}. In the former
reference the authors prove that, with a careful selection of the boundary
conditions, the variable of the Lax pair is time independent for (spatial)
$x = +\infty$, given initial conditions at $x = -\infty$. Thus a
Taylor expansion in the spectral parameter gives a series of non-local
conserved charges. In the latter paper, the authors apply the transformation
to the global (left, for instance) conserved current. This gives them a
spectral parameter dependant conserved current, which upon Taylor expansion
generates an infinite set of non-local currents, all of them conserved.

Although very well known and studied, in general and applied to this
particular model, the Lax pair has a somewhat obscure physical origin. Also,
there is no systematic way to find it in a given system. We hope to
contribute a small step to solve these problems. We show here a way to get
the Lax equation for the two dimensional $\sigma$-model, which somewhat
clarifies its physical meaning. This way of proceeding also enables the
study of a possible generalization to more dimensions.

We should stress here that a higher dimensional version of the non-local
symmetry of this model is sensible. Actually the (4D euclidean) self-dual
gauge theory has this kind of symmetry transformation \cite{Ch-Do}, which
is also related to several $(2+1)$-dimensional integrable systems
\cite{ChKeNe}. Moreover, this symmetry and its associated Lax pair is also
present in (at least) a subspace of the solution's manifold of a large class
of models based on vacuum general relativity \cite{Ga,BeSa}.

Aside from its own interest, the non-local symmetry of this system is
relevant in the Gribov problem. We must only realize that (classically) the
vacuum sector of the Yang-Mills theory is exactly the $n$-dimensional
$\sigma$-model if we choose the Lorentz ($\partial \cdot A = 0$) gauge. So
the existence of a continuous 1-parameter family of solutions of the
non-linear $\sigma$-model tells us of a continuum of vacua in the
non-abelian gauge theories in the Lorentz gauge.

\section{Constructing the Lax pair}

Only to fix our notation, we enumerate first some standard results
about this model. We write for the action
$$
S = \int {\d^d}x \, \tr [\partial_\mu(g^{-1}) \partial^\mu g] \ ,
$$
where $g$ is a ($G$) group valued field over Minkowsky space-time $M^d$.
This action has two global symmetries,
\begin{eqnarray*}
g(x) &\longmapsto & L \, g(x) \nonumber \\
g(x) &\longmapsto & g(x) R \ \ , \ \ \mbox{with}\ L,R \in G
                           \ \ \mbox{and independent of $x$} \ ,
\end{eqnarray*}
that tell us of the conservation of the (Noether) currents
$$
L_\mu \equiv g \, \partial_\mu g^{-1} \ \ \ \ \mbox{and}
           \ \ \ \ R_\mu \equiv g^{-1} \partial_\mu g \ ,
$$
which are related by $R_\mu = - g^{-1} L_\mu \, g$.

The equations of motion of this model can be written as\footnote{We will use
$\doteq$ for weak equality, {\em i.e.}, equality over field equations.}
\begin{equation}
     \mbox{$\sqcup\!\!\!\!\sqcap$} g +
            (\partial_\mu g \, \partial^\mu g^{-1}) g \doteq 0 \ ,
\label{f-eq}
\end{equation}
or as
\begin{equation}
\partial \cdot L \doteq 0
\label{divl}
\end{equation}
in terms of the left current, or equivalently $\partial \cdot R \doteq 0$ in
terms of the right one. These currents obviously satisfy a
``zero-curvature'' condition,
\begin{eqnarray}
F_{\mu\nu}(L) &\equiv & \partial_\mu L_\nu - \partial_\nu L_\mu
                     + [L_\mu, L_\nu] = 0                 \nonumber \\
F_{\mu\nu}(R) &\equiv & \partial_\mu R_\nu - \partial_\nu R_\mu
                     + [R_\mu, R_\nu] = 0 \ .
\label{fmunul}
\end{eqnarray}

With this notation, the linear (when left-multiplied by $U$) equations
\begin{equation}
U^{-1} \partial_\mu U = \1/2 \left[(1-\cosh (\lambda )) L_\mu -
               \sinh (\lambda ) \, \epsilon_{\mu\nu} L^\nu \right]
\label{lax-p}
\end{equation}
are called the Lax pair. $\lambda$ is the spectral parameter. The group
valued field $U$ should here be taken as the variable, for each given $L$.
It is straightforward to verify that the compatibility condition
$[\partial_\mu , \partial_\nu] U = 0$ is satisfied for pure gauge $L$'s with
zero divergence. In this case a formal solution to Eq.(\ref{lax-p}) is given
by
\begin{eqnarray*}
U(x) = U_o \, {\rm P} \exp \int \1/2
      \left[ (1-\cosh (\lambda )) L_\mu -
           \sinh (\lambda ) \, \epsilon_{\mu\nu}L^\nu \right] \d x^\mu \ ,
\end{eqnarray*}
where P is the path ordering operator (ordering from left to right), along
a path from a fixed point to $x$. A completely analogous Lax pair can be
constructed with the right current,
\begin{equation}
V^{-1} \partial_\mu V = \1/2
               \left[(1-\cosh (\lambda )) R_\mu -
               \sinh (\lambda ) \, \epsilon_{\mu\nu}R^\nu \right] \ .
\label{lax-p-v}
\end{equation}
B\"aklund transformations are at this point usually introduced.

We can in a sense reverse this argument, starting from a transformation
$g \longmapsto g'$ with
\begin{equation}
g'(x) = U(x) g(x) V^{-1}(x) \ ,
\label{tr}
\end{equation}
where $U$ and $V$ are some group valued fields.
This is a pretty general form, but looking at the action we realize we can
make it {\em a-priori\/} invariant asking for (\ref{tr}) to fulfill
\begin{equation}
\partial_\mu g' = \partial_\mu (U g V^{-1}) =
                  U \Lambda_{\mu\nu} \partial^\nu g \, V^{-1} \ .
\label{tr-cond}
\end{equation}
Here $\Lambda$ is an undetermined (eventually point dependant) Lorentz
transformation
\newline
($\Lambda_{\mu\nu}(x)\Lambda^{\mu\sigma}(x) = \delta_\nu^{\ \sigma}$).
This obviously makes the action invariant, but what is no so trivial to
fulfill is the consistency of condition (\ref{tr-cond}), that is $U$ and
$V$ such that this equation holds may not exist. Indeed in the
$(1+1)$-dimensional case this condition is only solvable over the field
equation solutions. Eq.(\ref{tr-cond}) can also be written as:
\begin{equation}
U^{-1} \partial_\mu U - g \, V^{-1} \partial_\mu V \, g^{-1}
                      = L_\mu - \Lambda_{\mu\nu} L^\nu \ ,
\label{tr-cond'}
\end{equation}
and implies for $L$ the transformation law
\begin{equation}
L'_\mu = U \Lambda_{\mu\nu} L^\nu U^{-1} \ .
\label{tr-L}
\end{equation}
We see that the Lax pairs (\ref{lax-p}) and (\ref{lax-p-v}) satisfy
Eq.(\ref{tr-cond'}) in the (1+1)-dimensional case, because in this case any
Lorentz transformation can be written as \
$\Lambda_{\mu\nu} = \cosh (\lambda ) \, g_{\mu\nu}
                    +\sinh (\lambda ) \, \epsilon_{\mu\nu} \,$. \
Anyway, Eq.(\ref{tr}) and Eq.(\ref{tr-cond'}) enable a testable path to
generalization to higher dimensions.

We can obtain the Lax pair doing a series of infinitesimal transformations,
with
$$
\Lambda_{\mu\nu} = g_{\mu\nu} + \omega_{\mu\nu} \ ,
                   \ \ \ \ |\omega_{\mu\nu}| \ll 1 \ ,
$$
where $\omega_{\mu\nu}$ is an antisymmetric tensor. For just one
transformation, we have
\begin{eqnarray}
U &=& 1 + u + {\cal O}(\omega_{\mu\nu}^2) \ , \nonumber \\
V &=& 1 + v + {\cal O}(\omega_{\mu\nu}^2) \ , \nonumber \\
U^{-1} \partial_\mu U &\simeq& \partial_\mu u \ , \ \ \ \
        V^{-1} \partial_\mu V \simeq \partial_\mu v \ ,
\label{it0}
\end{eqnarray}
where we have taken $U$ and $V$ so as to preserve boundary conditions for
$g$. That is, for the identity transformation we should have $g'=g$ so
$U(\omega_{\mu\nu} = 0) = V(\omega_{\mu\nu} = 0) = 1$.
Therefore we have
\begin{eqnarray}
L'_\mu &=& g' \partial_\mu {g'}^{-1}
        = (1+u) g (1-v) \, \partial_\mu [(1+v) g^{-1} (1-u)]    \nonumber \\
       &=& L_\mu - \partial_\mu u + [u, L_\mu]
           + g \, \partial_\mu v \, g^{-1} \ .
\label{it2}
\end{eqnarray}
But Eq.(\ref{tr-cond'}) gives
\begin{equation}
    \partial_\mu u = g \partial_\mu v g^{-1} - \omega_{\mu\nu} L^\nu \ ,
\label{it1}
\end{equation}
up to first order in $\omega$, so we get
\begin{equation}
L'_\mu = L_\mu + [u, L_\mu] + \omega_{\mu\nu} L^\nu
                            + {\cal O}(\omega^2) \ .
\label{it3}
\end{equation}

\noindent Calculating now the divergence of Eq.(\ref{it3}),
\begin{eqnarray}
\partial \cdot L' & \doteq & [\partial_\mu u, L^\mu]
         + \partial^\mu \omega_{\mu\nu} L^\nu               \nonumber \\
    &=& \left[ \partial_\mu u +
                \1/2 \omega_{\mu\nu} L^\nu , L^\mu \right] \ ,
\label{it4}
\end{eqnarray}
where constancy of $\omega_{\mu\nu}$ was assumed and Eq.(\ref{fmunul}) was
employed. If we impose $\partial \cdot L'$ to be zero (weakly, for all $L$
verifying Eq.(\ref{divl}) and Eq.(\ref{fmunul})), it is enough to take
\begin{equation}
\partial_\mu u + \1/2 \omega_{\mu\nu} L^\nu = \alpha L_\mu \ .
\label{it5}
\end{equation}

\noindent But for this equation to be consistent it should verify the
condition\footnote{Note that
$[\partial_\mu , \partial_\nu] u = F_{\mu\nu}(U^{-1} \partial U)$
up to leading order in $\omega$.}
\begin{equation}
[\partial_\mu , \partial_\nu] u = 0 \ ,
\label{it6}
\end{equation}
that is
\begin{equation}
     \partial_\nu \omega_{\mu\sigma} L^\sigma -
     \partial_\mu \omega_{\nu\sigma} L^\sigma +
     \alpha (\partial_\mu L_\nu - \partial_\nu L_\mu) = 0 \ .
\label{it7}
\end{equation}

If we now restrict ourselves to work in $1+1$ dimensions we can write
$\omega_{\mu\nu} = \gamma \epsilon_{\mu\nu}$. So we obtain from
Eq.(\ref{it7})
\begin{equation}
\alpha (\partial_\mu L_\nu - \partial_\nu L_\mu) = 0 \ ,
\label{it8}
\end{equation}
because for $\mu \neq \nu$ it is \
$\partial_\mu \epsilon_{\nu\sigma} L^\sigma -
 \partial_\nu \epsilon_{\mu\sigma} L^\sigma =
                 - \epsilon_{\mu\nu} \partial \cdot L \doteq 0 \,$. \
We can then choose $\alpha = 0$, so
$$
\partial_\mu u = -\1/2 \gamma \epsilon_{\mu\nu} L^\nu
$$
up to order $\gamma$. Renaming $\gamma$ as $\gamma_0$, we can now iterate
this procedure (obviously we must check consistency) doing a
transformation of parameter $\gamma_1$ to obtain $g_{(2)}$, $L_{(2)}^\mu$
starting from $g_{(1)}$, $L_{(1)}^\mu$, etc. This can be summarized as
shown below:

$$
\unitlength 1mm
\linethickness{0.4pt}
\begin{picture}(140.00,12.00)
\put(48.00,4.00){\vector(1,0){0.2}}
\put(25.00,4.00){\line(1,0){23.00}}
\put(93.00,4.00){\vector(1,0){0.2}}
\put(70.00,4.00){\line(1,0){23.00}}
\put(130.00,4.00){\vector(1,0){0.2}}
\put(115.00,4.00){\line(1,0){15.00}}
\put(13.00,4.00){\makebox(0,0)[lc]{$g$, $L$}}
\put(52.00,4.00){\makebox(0,0)[lc]{$g_{(1)}$, $L_{(1)}$}}
\put(97.00,4.00){\makebox(0,0)[lc]{$g_{(2)}$, $L_{(2)}$}}
\put(133.00,4.00){\makebox(0,0)[lc]{$\ldots$}}
\put(30.00,9.00){\makebox(0,0)[lc]{$\gamma_0$, $u_{(0)}$}}
\put(75.00,9.00){\makebox(0,0)[lc]{$\gamma_1$, $u_{(1)}$}}
\end{picture}
$$

Up to this point we have:
\begin{subequations}
\label{it9}
\begin{align}
\partial_\mu u_{(0)} &=
      -\1/2 \gamma_0 \epsilon_{\mu\nu} L_{(0)}^\nu \ , \\
L_{(1)}^\mu &= L_{(0)}^\mu + [u_{(0)}, L_{(0)}^\mu]
                + \gamma_0\epsilon^{\mu}_{\ \nu} L_{(0)}^\nu
                            + {\cal O}(\gamma_0^2) \ ,
\end{align}
\end{subequations}
and $v_{(0)}$ can be found from
$\partial_\mu u_{(0)} = g_{(0)} \partial_\mu v_{(0)} [g_{(0)}]^{-1}
                         - \gamma_0\epsilon_{\mu\nu} L_{(0)}^\nu$.
We want to show first that this procedure can be iterated as many times as
we want, and then we will show that this iteration leads us to a non
infinitesimal transformation, giving the usual Lax pair.

We started from a $L_{(0)}$ configuration with zero divergence and
curvature and, because of integrability of Eq.(\ref{it9}a), $L_{(1)}$
given by Eq.(\ref{it9}b) is well defined. To restart the process $L_{(1)}$
should also have zero divergence and curvature. But, by construction
$L_{(1)}$ has zero curvature up to leading order in $\gamma_0$, as it has
been defined as \
$L_{(1)}^\mu = g_{(1)} \partial^\mu g_{(1)}^{-1}$\ ,
neglecting higher order terms. $L_{(1)}$ has also zero divergence, as
$u_{(0)}$ was chosen for this purpose. Taking into account the discussion in
the Appendix a well defined procedure for integrating the analog of
Eq.(\ref{it9}a) for the second iteration can be given.

With exactly the same reasoning we see that for any $n$
\begin{subequations}
\label{it10}
\begin{align}
\partial_\mu u_{(n)} &=
      -\1/2 \gamma_n \epsilon_{\mu\nu} L_{(n)}^\nu \ , \\
L_{(n+1)}^\mu &= L_{(n)}^\mu + [u_{(n)}, L_{(n)}^\mu]
                + \gamma_n\epsilon^{\mu}_{\ \nu} L_{(n)}^\nu
\end{align}
\end{subequations}
provide us with\footnote{
This can also be verified inductively. This way it is easier to see that:
\begin{itemize}
\item[-]{ $[\partial_\mu, \partial_\nu] u_{(n)} =
          \1/2 \gamma_n \epsilon_{\mu\nu} \partial \cdot L_{(n)} =
          \gamma_n {\cal O}(\gamma_{n-2}^2, \ldots , \gamma_0^2)$}
\item[-]{ $\partial \cdot L_{(n+1)} = \partial \cdot L_{(n)} +
             [u_{(n)}, \partial \cdot L_{(n)}] - \gamma_n F_{01}(L_{(n)}) =
                    {\cal O}(\gamma_{n-1}^2, \ldots , \gamma_0^2)$}
\item[-]{ $F_{\mu\nu}(L_{(n+1)}) = F_{\mu\nu}(L_{(n)}) -
           \gamma_n \epsilon_{\mu\nu} \partial \cdot L_{(n)} +
           [u_{(n)}, F_{\mu\nu}(L_{(n)})] $ \\
       \hspace*{2.2cm} $ + (\mbox{terms with $\gamma_n u_{(n)}$,
           $\gamma_n^2$, $u_{(n)}^2$, all are ${\cal O}(\gamma_n^2)$}) $ \\
       \hspace*{1.9cm} $ = {\cal O}(\gamma_n^2, \ldots , \gamma_0^2) $}
\end{itemize}
}:
\begin{itemize}
\item[(I)]{ a well defined $L_{(n+1)}^\mu$\ , because
     $[\partial_\mu, \partial_\nu] u_{(n)} =
         \1/2 \gamma_n \epsilon_{\mu\nu} \partial \cdot L_{(n)}$}
\item[(II)]{ a zero divergence $L_{(n+1)}^\mu$\ , up to order
             $\gamma_{n-1}$}
\item[(III)]{ a zero curvature $L_{(n+1)}^\mu$\ , up to order
              $\gamma_n$.}
\end{itemize}

We now want to find $L_{(n)}^\mu$ and $g_{(n)}$ explicitly in terms of
$L^\mu \equiv L_{(0)}^\mu$ and $g \equiv g_{(0)}$. Following
Eq.(\ref{tr}) and Eq.(\ref{it0}), $g_{(n)}$ is given by
\begin{eqnarray*}
g_{(n+1)} &=& (1+u_{(n)}) g_{(n)} (1-v_{(n)}) \nonumber \\
          &=& (1+u_{(n)}) (1+u_{(n-1)}) \ldots (1+u_{(0)}) g_{(0)}
            (1-v_{(0)}) \ldots (1-v_{(n)}) \ .
\label{it11}
\end{eqnarray*}
What we need is $U = \mbox{ (Limit of) }(1+u_{(n)}) \ldots (1+u_{(0)})$, or
considering the Lax pair we want to find, the equation satisfied by $U$: \
$U^{-1} \partial_\mu U = \mbox{(something)}$. For $n=0$ we have
\begin{eqnarray*}
U_0^{-1} \partial_\mu U_0 &=&
            (1-u_{(0)})\partial_\mu (1+u_{(0)})             \nonumber \\
            &=& \partial_\mu u_{(0)} + {\cal O}(\gamma_0^2) =
             -\1/2 \gamma_0 \epsilon_{\mu\nu} L_{(0)}^\nu \ .
\label{it12}
\end{eqnarray*}

\noindent For an arbitrary $n$ we get\footnote{
We write $\prod_{i=0}^n a_{i}$ for $a_0 \ldots a_n$, and
$\prod_{i=n}^0 a_{i}$ for $a_n \ldots a_0$. }
\begin{eqnarray*}
U_n^{-1} \partial_\mu U_n &=&
      \prod_{i=0}^n (1-u_{(i)}) \: \partial_\mu \prod_{i=n}^0 (1+u_{(i)})
                                                            \nonumber \\
	   &=& \sum_{j=0}^n \left[ \prod_{i=0}^{j-1} (1-u_{(i)})
            \: \partial_\mu u_{(j)} \prod_{i=j-1}^0 (1+u_{(i)})
            + {\cal O}(\gamma_n^2, \ldots, \gamma_j^2) \right] \ .
\label{it13}
\end{eqnarray*}

\noindent Here products $(1-u_{(i)}) (1+u_{(i)})$, coming from the terms in
$U^{-1}$ and $\partial_\mu U$ respectively, cancel to first order in the
parameter. If we now rewrite Eq.(\ref{it10}b) as
\begin{eqnarray}
L_{(i)}^\mu &=& L_{(i-1)}^\mu + [u_{(i-1)}, L_{(i-1)}^\mu]
           + \gamma_{i-1}\epsilon^{\mu}_{\ \nu} L_{(i-1)}^\nu \nonumber \\
            &=& (1+u_{(i-1)})
                 (g^\mu_{\ \nu} + \gamma_{i-1} \epsilon^{\mu}_{\ \nu})
               L_{(i-1)}^\nu (1-u_{(i-1)}) \ ,
\label{it14}
\end{eqnarray}
up to higher order terms, and we use Eq.(\ref{it10}a), we get
\begin{eqnarray*}
U_n^{-1} \partial_\mu U_n &=&
	   \sum_{j=0}^n -\1/2 \gamma_j \epsilon_{\mu{\nu_j}} \left[
	   \prod_{i=0}^{j-2} (1-u_{(i)}) \:
    (g^{\nu_j}_{\ \nu_{j-1}} + \gamma_{j-1} \epsilon^{\nu_j}_{\ \nu_{j-1}})
    L_{(j-1)}^{\nu_{j-1}} \prod_{i=j-2}^0 (1+u_{(i)}) \right] \nonumber \\
      & & + {\cal O}(\gamma_n^2, \ldots, \gamma_{0}^2) \ .
\label{it15}
\end{eqnarray*}
In this last equation factors $(1-u_{(j-1)})$ and $(1+u_{(j-1)})$ cancel
at each side of the $L_{(j-1)}$. It is now easy to see that after employing
Eq.(\ref{it14}) $\; (j-1)$ times more we obtain
\begin{eqnarray*}
U_n^{-1} \partial_\mu U_n &=&
      \sum_{j=0}^n -\1/2 \gamma_j \epsilon_{\mu{\nu_j}}
   (g^{\nu_j}_{\ \nu_{j-1}} + \gamma_{j-1} \epsilon^{\nu_j}_{\ \nu_{j-1}})
   \ldots (g^{\nu_1}_{\ \nu_{0}} + \gamma_{0} \epsilon^{\nu_1}_{\ \nu_{0}})
      L_{(0)}^{\nu_{0}} \ .
\label{it16}
\end{eqnarray*}

\noindent Expanding the sums and products, the final expression for this is
\begin{eqnarray*}
U_n^{-1} \partial^\mu U_n =
     - \1/2 \left( \sum \gamma + \sum_{\neq}\gamma\gamma\gamma
                   + \ldots \right) \epsilon^\mu_{\ \nu} L_{(0)}^\nu
     - \1/2 \left( \sum_{\neq} \gamma\gamma
                   + \ldots \right) L_{(0)}^\mu \ ,
\label{it17}
\end{eqnarray*}
where all sums are with indices ranging from $0$ to $n$, all indices are
different within the sums and each term is included only once, {\em i.e.},
$(\sum_{\neq} \gamma\gamma )$ means $(\sum_{i_1 = 0}^n \sum_{i_2 > i_1}^n
\gamma_{i_1} \gamma_{i_2})$, etc.

\noindent If we now fix all parameters $\gamma_i$ to be equal, and define
them as $\lambda / (n+1)$, we obtain
\begin{eqnarray*}
  & & \sum \gamma = \sum_{i=0}^n \gamma_i = \lambda  \ \ \ , \nonumber \\
  & & \sum_{\neq} \gamma\gamma = \sum_{i=0}^n \sum_{j>i}^n \gamma_i \gamma_j
	 = \frac{\lambda^2}{(n+1)^2}
            \left( \begin{matrix}
                          {N+1} \\ 2
                   \end{matrix} \right)
	 = \frac{\lambda^2}{2!} + {\cal O}\left( 1/n \right)
                                                   \ \ \ , \nonumber \\
  & & \sum_{\neq}\gamma\gamma\gamma = \frac{1}{3!} \lambda^3
         + {\cal O}\left( 1/n \right) \ \ \ ,        \nonumber \\
  & & \mbox{etc.} \ ,
\label{it18}
\end{eqnarray*}
whenever the order of the term is much lower than $n$. Therefore we get, in
the large $n$ limit, a Taylor series. Obviously this series adds up to
\begin{equation}
U^{-1} \partial^\mu U =
       \1/2 \left[ \left( 1 -\cosh(\lambda) \right) L^\mu
                  - \sinh(\lambda) \; \epsilon^{\mu\nu} L_\nu \right] \ ,
\label{it19}
\end{equation}
that is what we had in Eq.(\ref{lax-p}).

\section{Discussion and conclusions}

We have obtained the Lax pair for the non-linear $\sigma$-model in a mainly
constructive way. It can now be given a physical significance, because it is
the compatibility condition for the existence of a family of
(non-linear/non-local) weak symmetry transformations for the model. The
spectral parameter of such a family appears here as the parameter of a
Lorentz transformation involved in the invariance of the action under the
symmetry transformation. Built this way, it seems more realistic to think
about its possible generalization to higher dimensions. Anyway, it is not an
easy thing: for a $(2+1)$-dimensional space-time the construction breaks
down at Eq.(\ref{it8}) because the first two terms of Eq.(\ref{it7}) do not
cancel anymore, so $\alpha$ can not be chosen as zero to make Eq.(\ref{it6})
hold. One can try to restrict the symmetry to only a subset of the
solutions' space, considering for instance only those fields $g$ that are
axially symmetric. Since we want to iterate the transformation in such a way
that the new solution has also axial symmetry, the problem is that for
physical consistency the $\omega_{\mu\nu}$ tensor of parameters should also
be axially symmetric. This means that the (cartesian) components
$\omega_{\mu\nu}$ are position dependant. Therefore, if we want a
non-trivial transformation, the second part of Eq.(\ref{it4}) must be
modified with the addition of a term $ L^\nu \partial^\mu \omega_{\mu\nu}$.

\newpage
\appendix
\section*{Appendix: Path independence}
\label{s-path-indep}

Note that being
$[\partial_\mu, \partial_\nu] u_{(n)} = 0 + {\cal O}(\gamma_i^2)$
({\em i.e.}, not exactly zero in general) we still have a ``small'' path
dependency in $u_{(n)}$. How can we deal with it?

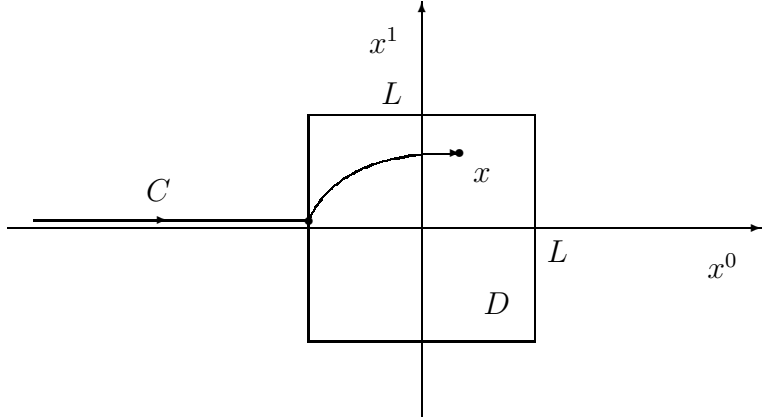
\begin{figure}
\unitlength 1mm
\linethickness{0.4pt}
\begin{picture}(120.00,65.00)(0.00,8.00)
\put(75.00,65.00){\vector(0,1){0.2}}
\put(75.00,10.00){\line(0,1){55.00}}
\put(120.00,35.00){\vector(1,0){0.2}}
\put(20.00,35.00){\line(1,0){100.00}}
\put(60.00,20.00){\framebox(30.00,30.00)[cc]{}}
\put(115.00,30.00){\makebox(0,0)[cc]{$x^0$}}
\put(70.00,60.00){\makebox(0,0)[cc]{$x^1$}}
\put(23.33,36.00){\line(1,0){36.67}}
\put(41.00,36.00){\vector(1,0){0.2}}
\put(60.00,36.00){\circle*{1.00}}
\put(80.00,45.00){\vector(1,0){0.2}}
\multiput(60.00,36.00)(0.11,0.21){8}{\line(0,1){0.21}}
\multiput(60.88,37.71)(0.11,0.15){10}{\line(0,1){0.15}}
\multiput(61.98,39.23)(0.11,0.11){12}{\line(0,1){0.11}}
\multiput(63.31,40.58)(0.15,0.12){10}{\line(1,0){0.15}}
\multiput(64.85,41.75)(0.20,0.11){9}{\line(1,0){0.20}}
\multiput(66.62,42.74)(0.28,0.12){7}{\line(1,0){0.28}}
\multiput(68.61,43.55)(0.37,0.11){6}{\line(1,0){0.37}}
\multiput(70.82,44.18)(0.61,0.11){4}{\line(1,0){0.61}}
\multiput(73.25,44.64)(0.89,0.09){3}{\line(1,0){0.89}}
\put(75.91,44.91){\line(1,0){4.09}}
\put(80.00,45.00){\circle*{1.00}}
\put(83.00,42.00){\makebox(0,0)[cc]{$x$}}
\put(93.00,32.00){\makebox(0,0)[cc]{$L$}}
\put(71.00,53.00){\makebox(0,0)[cc]{$L$}}
\put(85.00,25.00){\makebox(0,0)[cc]{$D$}}
\put(40.00,40.00){\makebox(0,0)[cc]{$C$}}
\end{picture}
\caption{Setup for integration of Eq.(\ref{it10}a)}
\label{fig2}
\end{figure}

We aim at obtaining a (``true'') function $g'(x)$, starting from $g(x)$,
making $N$ steps of parameters $\gamma_0, \dots , \gamma_{N-1}$, and finally
taking the limit $N \to \infty$, with (for all $i$) $N \gamma_i = \lambda$,
a non infinitesimal quantity. However, before the limit procedure, the
$N$-times iterated ``function'' $g_{(N)}(x)$ may be defined only on a
bounded domain $D_N$ of $M^2$. We can choose, for instance, a square region
$|x^\mu| < L$, with $L^2 \sim 1/\max(\gamma_i)$. So for $x \in D_N$ we can
find $u_{(i)}(x)$ from Eq.(\ref{it10}a) integrating over a path $C$,
consisting of a fixed piece from $-\infty$ to some point on the boundary of
$D_N$, and an arbitrary piece {\em inside} $D_N$ to $x$ (as shown in
Fig.\ref{fig2}). Therefore, integrating over two different paths we get
two different $u$'s, with
\begin{eqnarray*}
|\Delta u_{(i)}| &\sim &
         |[\partial_0 , \partial_1] u_{(i)}| |\mbox{Area between paths}|
											      \nonumber \\
     &\leq & |\gamma_i| |\partial \cdot L_{(i)}| \, L^2
     \leq \frac{\gamma_i}{\max(\gamma_j)} \, \gamma_{i-2}^2 \ ,
\end{eqnarray*}
and we obtain a ``fixed'' error for $g_{(N)}(x)$ whenever $x$ is inside
$D_N$, with a given $\lambda = N \gamma_i$.

\newpage

\end{document}